\documentclass{eas}
\usepackage{graphicx}
\def\etal{{\it et al.\ }}
\def\msun{{\rm\,M_\odot}}

\newcommand\aj{AJ}

\newcommand\apj{ApJ}
\newcommand\apjl{ApJ}
\newcommand\apjs{ApJS}
\newcommand\aap{A\&A}
\newcommand\mnras{MNRAS}

\newcommand\nat{Nature}
\newcommand\apss{Ap\&SS}

\begin{document}

\title{The role of the Interstellar Medium in Galaxy Formation Simulations} 
\runningtitle{The ISM in Galaxy Formation}
\author{Greg L. Bryan}\address{Department of Astronomy, Columbia University, 550 West 120th Street, New York, NY 10027, USA}
%
%
\begin{abstract}
There is a strong connection between the formation of a disk galaxy and the properties of the interstellar medium (ISM).  Theoretical work has typically either focused on the cosmological buildup of a galaxy with a relatively crude model for the gas physics, or examined local processes in the ISM and ignored the global evolution of the galaxy itself. Here, I briefly review what has been learned from both of these approaches, and what can be done to bridge the gap between them.  I argue that cosmological simulations need to learn from observational and theoretical work on local ISM properties and adopt more sophisticated models for the processes that they cannot resolve.  Since the ISM is still incompletely understood, there are a number of reasonable approaches for these ``subgrid" models, and I will discuss the strengths and limitations of each.
\end{abstract}
\maketitle

\section{Introduction}

Galaxies are one of the fundamental building block of classical astronomy -- they contain the vast majority of stars in the universe and their large-scale distribution in space, as observed by redshift surveys, serves to constrain key cosmological parameters.  However, their formation --- and even to some extent their constituents --- are still only poorly understood.  While modern cosmological theories can reliably predict the distribution of dark matter halos, the distribution of the gas and stars inside is more difficult.  Unfortunately, this distribution is crucial in order to make comparisons between, for example, observed and predicted galaxy rotation curves.

The large-scale distribution of galaxies, as probed by redshift surveys (e.g., Eisenstein \etal 2005), provides an excellent match between observations and the cosmological constant-dominated cold-dark matter model ($\Lambda$CDM).  These comparisons are based on high-resolution N-body simulations involving only dark matter.  However, such simulations generally do not resolve the structure of galaxy-sized halos with any detail.  More computationally intensive simulations of single galaxy halos have been performed (Power \etal 2003; Moore \etal 2004; Hayashi \etal 2004; Navarro \etal 2004).  While the details are still somewhat uncertain, these simulations and others generally predict a strongly cusped central density profile, with the dark matter density increasing as $r^{-1}$ or steeper (Diemend \etal 2004; Navarro \etal 2004; Fukushige \etal 2004).  Such a steep inner dark matter core has been claimed to be in conflict with the observed velocity profiles of low-surface brightness galaxies and dark-matter dominated dwarf systems (de Blok \etal 2003, 2004), although the impact of observational uncertainties is still controversial (van den Bosch \etal 2000; Swaters \etal 2003).  In addition to this ``cusp" problem, it has been noted that the number of dark matter subhalos in galactic-size halos is much larger than the observed number of galactic satellites (e.g., Klypin \etal 1999; Moore \etal 2000).

These comparisons are in some sense between apples and oranges, as the simulations predict the distribution of dark matter without taking into account the gas physics.  Even in galaxies which are currently dark-matter dominated, the observables come from the gas distribution: generally either 21 cm emission from HI or H$\alpha$ emission (e.g. Chung \etal 2002; de Blok \etal 2004).  This leads us to consider simulations that include gas dynamics and star formation.  In this review we restrict ourselves largely to disk galaxies, as this is where ISM physics play the largest role.

\subsection{Problems in Cosmological Disk formation}

The basic picture of disk formation in a dark-matter dominated cosmology was laid out by Fall \& Efstathiou (1980).  In this model, the proto-galactic gas and dark matter cloud is tidally torqued by large-scale density inhomogeneities, resulting in a net angular momentum.  The dark matter angular momentum is largely unimportant for the support of the halo, as it virializes before rotation dominates support.  On the other hand the gas can shock heat and radiate its binding energy, allowing further collapse beyond the virialization point.  This collapse is ultimately halted by the angular momentum of the gas, which is assumed to be preserved during the collapse.  For a dark matter halo, we can parameterize its total angular momentum $J$ in terms of the spin parameter
$$
\lambda = \frac{J |E|^{1/2}}{GM^{5/2}}
$$
where $E$ is the total energy and $M$ the total mass.  Simulations indicate typical values of $\lambda \sim 0.05$, and if the gas collapses within the dark-matter dominated halo, then rotation dominates when $r_{\rm disk} \approx \lambda r_{\rm vir}$.  Since the typical virial radius of galactic-sized halo is a few hundred kpc, this results in approximately the correct disk size.

Early cosmological simulations of galaxy formation (Navarro \& Benz 1991; Navarro \& White 1994; Steinmetz \& M\"uller 1995) found that stars and gas lost angular momentum due to dynamical friction during the early collapse of the disk.  This resulted in disks which were much smaller than observed. 

More recent simulations (e.g., Abadi et al 2003a,b; Okamoto \etal 2005), with improved resolution, have found that a rotationally-supported gas disk does form with most of the correct characteristics (including a reasonable scale-radius), but that an overly-massive spheroidal component creates a circular velocity profile that {\it rises} towards the center.  

The essential difficulty seems to be that during the multiple mergers that occur during the early stages of galaxy formation, dense knots of gas are formed.  These small concentrated clumps lose angular momentum, probably via dynamical friction, and spiral into the centers of the halos.  These gas clumps are identified with the too large spheroid component seen at $z=0$.  This interpretation is boosted by the observation that more realistic disks are found in simulations without small-scale structures and merging, such as in warm dark matter models (e.g. Sommer-Larsen \etal 2003).

Since the warm dark matter mass required to resolve the discrepancy seems to be currently at odds with other measurements such as the Lyman-alpha forest statistics (Viel \etal 2005), the real problem seems to be the formation of overly dense knots, almost surely due to an inadequate modeling of feedback in the ISM.  It is for this reason that we focus this review on ISM models.

Simulations which include some sort of ``sub-grid'' model to enhance the efficiency of feedback (Weil, Eke \& Efstathiou 1998; Thacker \& Couchman 2000; Sommer-Larsen \etal 2003; Brook \etal 2004; Robertson \etal 2004) tend to suppress the formation of an overly-massive spheroid.  This is promising, but there are a wide variety of such models in the literature and it is not clear if they are consistent with our current understanding of the Milky Way's own ISM -- this topic is addressed in the rest of the review (sections~\ref{sec:subgrid_models} and \ref{sec:ism_models}), but first we examine the possibility that the issue is entirely numerical, and in particular if it might be due to the way in which these simulations are carried out.  This is the topic of the next section.

However, first we mention briefly a related problems in galaxy formation.  This is the low fraction of baryons observed in galactic-sized halos.  For example, using weak-lensing to measure the total halo mass, Mandelbaum \etal (2006) find that the observed baryon fraction of Milky-Way like halos is approximately 20\%, compared to the 80\% of stars and gas that Abadi \etal (2003) find in their galactic disk.  It is not known if these baryons reside in some form of hot phase inside the galactic halo, or if they have been entirely expelled from their original halo.


\section{Adaptive Mesh Refinement Galaxy Formation Simulations}

To date, essentially all simulations of galaxy formation have been carried out with Smoothed Particle Hydrodynamics (SPH), due to its Lagrangean nature: resolution elements naturally follow the mass. This contrasts with Eulerian methods (which are preferred outside of astrophysics) which generally have a static mesh.  As a diffuse proto-galactic cloud collapses to small scales, its typical length scale approaches the mesh-spacing and the method breaks down.  Newly developed adaptive mesh approaches (e.g., Gnedin 1996; Bryan 1999; Fryxell 2000) allow the mesh to refine itself to match the solution, so that the galaxy is well resolved at all times.

In this section, I examine the application of the adaptive mesh refinement {\it Enzo} code to the problem of galaxy formation.  The {\it Enzo} code, described in Bryan \& Norman (1997), Norman \& Bryan (1999), Bryan (1999) and O'Shea \etal (2005), is a three-dimensional hydrodynamics code that uses a grid-based scheme for the gas and a particle-mesh method for the stars.  A large advantage to using AMR codes over static grids is that individual regions of the simulation box can have different levels of refinement. This significantly reduces computational time by only refining areas that need it. 

The AMR technique has shown some success in resolving the multi-phase nature of the interstellar medium in the presence of star formation and feedback (e.g., de Avillez \& Breitschwerdt 2004, Slyz \etal 2005). Particle-based codes like SPH tend to smooth the hot and cold phases which leads to an over estimate of the cooling rate (although improvements to the basic SPH method have been suggested: e.g., Marri \& White 2003; Springel \& Hernquist 2003).  It has been suggested that grid codes suffer less from this problem as the mesh allows sharp boundaries to be better resolved. 

\begin{figure}
\begin{center}
\includegraphics[width=5.5cm]{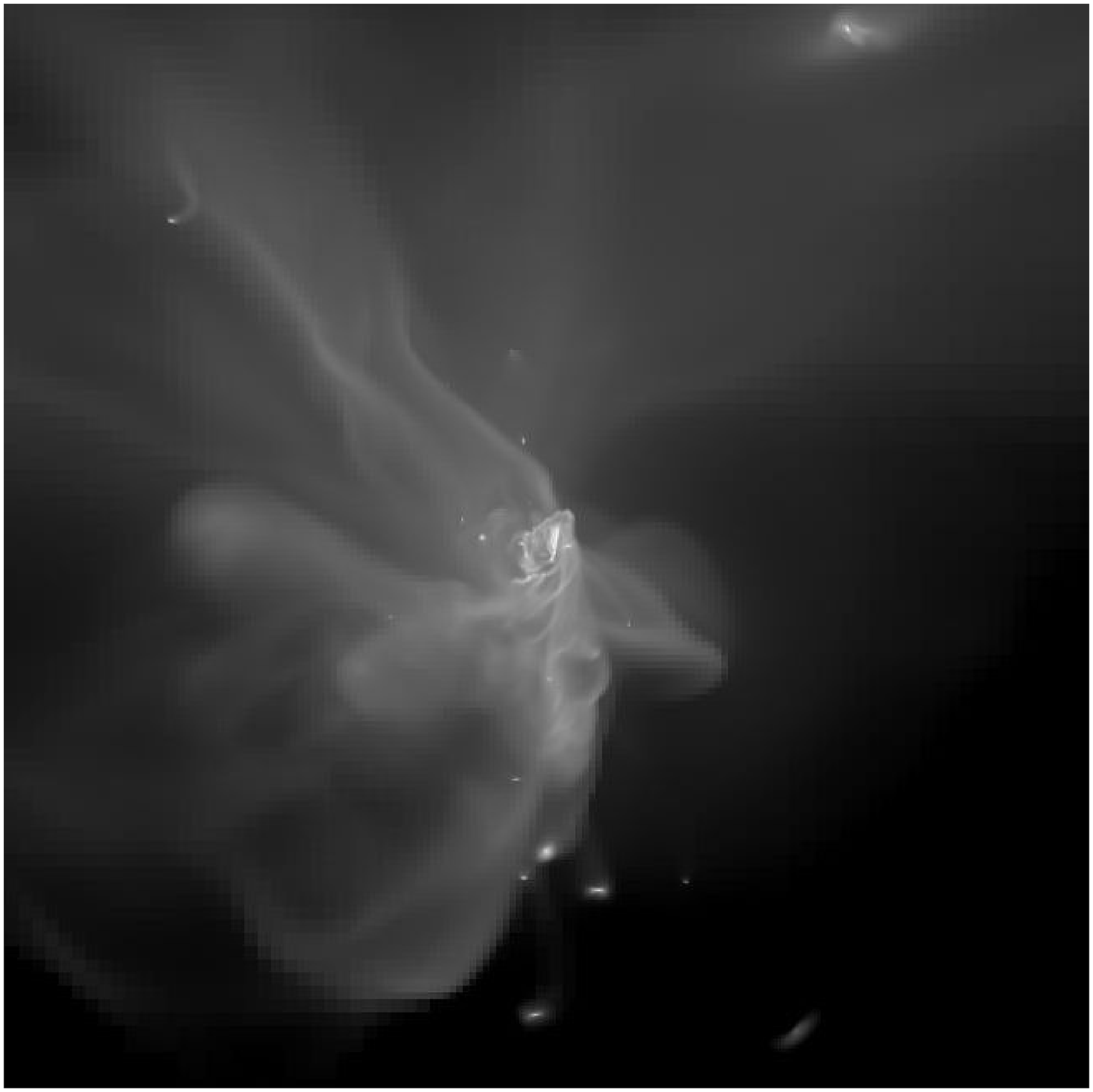}
\includegraphics[width=5.5cm]{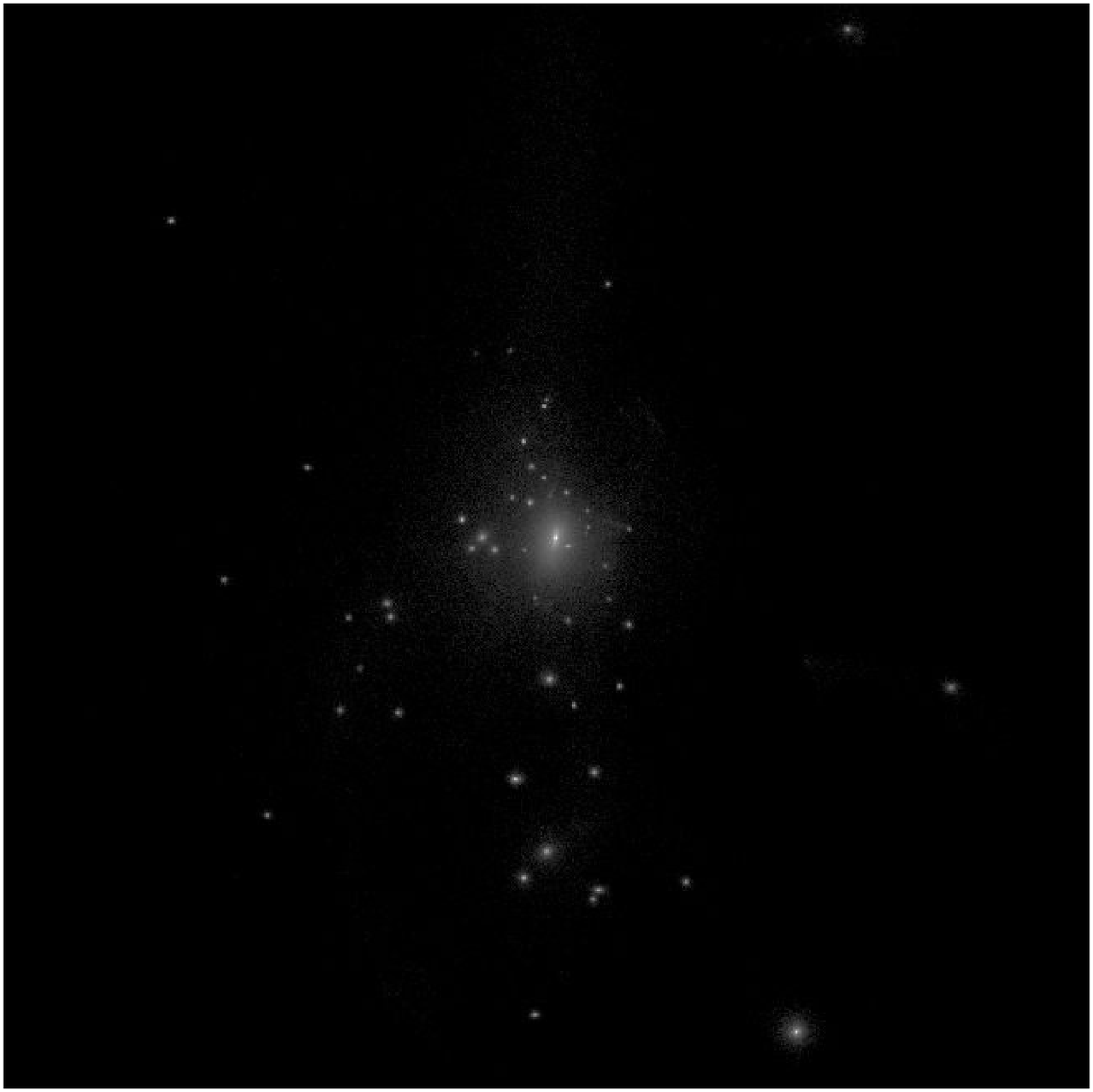}
\caption{This figure shows the projected gas density (left panel) and stellar density (right panel) from an adaptive-mesh refinement galaxy formation simulation. The images depict a region 500 comoving kpc on a side at $z=0.9$.}
\label{fig:galaxy_dm_stars}
\end{center} 
\end{figure}

\begin{figure}
\begin{center}
\includegraphics[height=6cm]{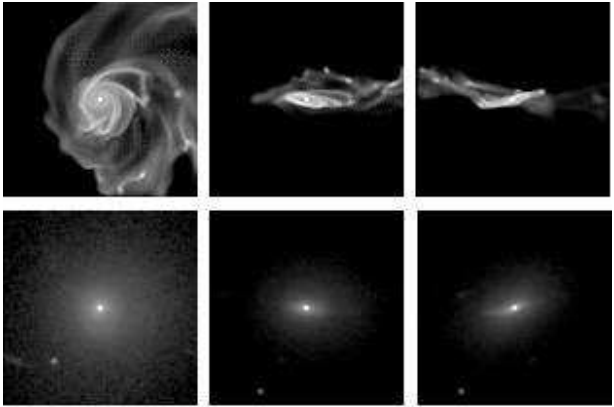}
\caption{These images (about 50 kpc across) show a close-up of the simulated disk rotated so that it is face-on, and edge-on.  The top set of images show the gas in the disk, while the bottom images depict the stellar density distribution. }
\label{fig:amr_disk}
\end{center}
\end{figure}

Figures \ref{fig:galaxy_dm_stars} and \ref{fig:amr_disk} show the gas and stellar mass distribution from such an AMR simulation.  The simulation is performed in a periodic box 15 Mpc on a side (this box is too small to be a fair representation of the universe, but this is not expected to strongly affect the disk properties), run in a $\Lambda$CDM universe with parameters consistent with current observations.  The dark matter particle mass is $9 \times 10^5 \msun$, and the resolution in the gas is of order $10^5 \msun$.  There are nearly a million dark matter particles within the virial radius. The cell size on the most refined level is about 200 comoving pc.

In these figures, I have focused in on a $8 \times 10^{11} \msun$ dark matter halo, demonstrating a highly disturbed disk in the center.  On closer examination (see Figure~\ref{fig:amr_disk}), the disk shows a warped morphology, probably in part due to a number of recent mergers (gas remnants of some of the merging objects can be seen as filamentary structures surrounding the galaxy).  An examination of the dark matter distribution (not shown here) demonstrates that stars inhabit primarily the centers of nearly all the dark matter clumps.  There are approximately 60 stellar satellites, which is (probably) considerably more than the observed Milky Way satellite population, a demonstration of the so-called galactic sub-structure problem (Klypin \etal 1999; Moore \etal 1999).  

Already by $z=0.9$, nearly 70\% of the gas has been converted to stars, and a close examination shows that most of these stars have ended up in a central spheroid.  Although this simulation has not been, so far, run past $z=0.9$, based on lower-resolution simulations (which show very similar results), we expect a stellar disk to form out of the gas disk, but as described in the introduction, an overly-massive spheroidal component has formed.  This can be seen in close-ups of the disk, shown in Fig.~\ref{fig:amr_disk}, which clearly shows both a huge, dense bulge, as well as a much more extended gas disk.

\begin{figure}
\begin{center}
\includegraphics[height=5.5cm]{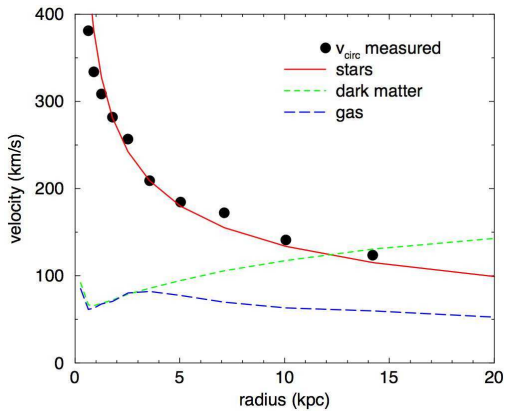}
\caption{The circular rotation curve for the simulated galaxy at $z=0.9$.  The solid points show the measured circular velocity of the gas in the simulation, while the lines show the contribution ($\sqrt(GM(r)/r)$) from various components.  At large radius, the gas disk is warped, biasing the measured gas circular velocity to low values.}
\label{fig:vcirc}
\end{center} 
\end{figure}

The massive spheroid has a strong impact on the circular velocity profile, as shown in Figure~\ref{fig:vcirc}.  The rotational velocity is peaked in the center of the galaxy to unrealistically large values and actually falls at larger radii.  This figure, which looks nearly identical to a figure in Abadi \etal (2003), demonstrates agreement between AMR and SPH results, although we have not yet carried out a systematic resolution study or simulated enough halos for a firm conclusion to be drawn.  It argues (but does not prove) that, while numerical problems associated with angular momentum transport are clearly present (e.g. Kaufmann \etal 2006a), they are not the primary source of the disagreement between observations and simulations.

\section{Subgrid models in Cosmological Simulations}
\label{sec:subgrid_models}

Since it seems unlikely that the problem is due entirely to numerical problems, we turn to an examine of the physics used to simulate the ISM.  Since none of these simulations can hope to resolve the physics of the interstellar medium (which operates on scales 100 to 1000 times smaller than typical disk scales), it is inevitable that some ``subgrid" model of the ISM must be used.  Here we classify these models into five different general types:

\subsection{Isothermal models.}  In galaxy collision simulations, it is often assumed that the ISM can be modeled by an isothermal equation of state (e.g. Barnes \& Hernquist 1992; Li, Mac Low \& Klessen 2005).  While this obviously ignores the multi-phase structure of the ISM, it does effectively simulate a non-thermal pressure component which is assumed to dominate for temperatures below some critical ``effective" temperature or sound-speed, which is often taken to be around $10^4$ K.

\subsection{Minimum Temperature models.} The isothermal model can be generalized to one in which an energy equation with heating and cooling terms is used.  This allows for shock-heating and radiative cooling, and so is clearly more realistic in that it can capture the hot phase of the ISM.  However, if cooling is allowed to low temperatures, then the gas will quickly cool to the bottom of the allowed cooling curve.  Again, in order to mimic processes which are not resolved (such as a multi-phase medium or turbulence), an artificially large minimum temperature is used.  Typically this is $10^4$ K (Steinmetz \& Navarro 1994, Abadi \etal 2003, and in the AMR simulation described earlier), but some works have advocated even larger values, around $3 \times 10^4$ K (Kaufmann \etal 2006b).

These minimum-temperature models have been criticized in Robertson \etal (2004) and Springel, Di Matteo \& Hernquist (2005) on the basis that the disk is highly Toomre unstable and will fragment into small clumps that lose angular momentum due to dynamical friction.  However, both Li, Mac Low \& Klesson 2005 and Kaufmann \etal 2006a,b have argued that these simulations are under-resolved and either the disks are not unstable, or that the fragments that do form in well-resolved simulations are not large enough to lose angular momentum in this way.  While this issue remains still somewhat open, no (published, $z=0$) minimum temperature model has successfully produced a galactic disk in a full cosmological simulation.

\subsection{Cooling Suppression models.}

One of the outstanding problems in galaxy formation is how to model the energetic input from supernovae and massive stars (not to mention AGN).  The problem is that, at the resolution that can be obtained in cosmological simulations (typically at best a few hundred pc), the blast wave from even a large number of nearly simultaneous supernovae is not resolved.  This means that if, as is usually the case, the feedback is modeled by injecting the energy into the local thermal energy of a particle or cell, it is spread over an artifically large region.  This results in the shocked gas being cooler and denser than it should be (i.e. than in the post-shock region of a SN bubble), and the energy is quickly radiated away.

One way around this problem is to simply suppress radiative cooling in the vicinity of a burst of star formation (which is followed quickly by the feedback from the short-lived massive stars).  This was first suggested in Gerritsen (1997) and is the basis for many of the latest (and most successful) attempts to produce realistic galactic disks (Thacker \& Couchman 2001; Brook \etal 2004, Stinson \etal 2006).  Typically this is done by turning off radiative cooling in particles within a resolution element of a newly-formed star particle for a few million years (see Governato \etal 2006 for a somewhat more involved prescription).

The result -- in for example Governato \etal (2006) -- is a galactic disk with a reasonable size, and disk characteristics such that it falls on the Tully-Fisher relation between luminosity and rotational velocity.  On the other hand, only a few models have been computed and the rotation curves are still not entirely realistic, so more work is clearly warranted.  Finally, the physical mechanism behind this approach is difficult to justify in detail; for example models of the local ISM do not find that radiative cooling is negligible in post-SN regions.

\subsection{High-efficiency or self-propagating star formation models.}

A related approach is to make the feedback so efficient that it overwhelms cooling, at least for a short time.  In Sommer-Larsen \etal (2003) and related work, star formation has been modeled with a redshift-dependent efficiency and a self-propagating component that triggers more star formation.  This makes feedback particularly efficient at high-redshift and seems to also result in realistic looking galactic disks.

\subsection{Multi-phase models.}

All of the previous approaches have made changes to the hydrodynamics algorithm that effectively mimic a ``sub-grid" model.   Another approach is to build in an explicit sub-grid model, so that there are additional variables that describe the state of the gas on scales smaller than can be resolved. 

One popular example of this is to implement a multi-phase description of the gas, such that each fluid element has a single pressure but can consist of gas clumps with a variety of densities and temperatures (each in pressure balance) on scales smaller than the smoothing-length, or cell size, of the simulation.  A large number of such methods have been implemented (e.g., Yepes \etal 1997, Semelin \& Combes 2002), but here we focus on the mostly widely used version, that of Springel \& Hernquist (2003).  In their implementation, radiative cooling serves to transfer matter from the hot to the cold phase, while heat conduction and feedback drive gas in the opposite direction.  They find that the description can be further simplified by assuming that, above some minimum density, there is a local balance between heating and cooling, so that the resulting pressure depends only on the density.  This polytropic-like relation is steeper than isothermal so that the effective temperature at a density of 1 cm$^{-3}$ is nearly $10^5$ K.  Because of the monotonic relation between density and pressure, this equation of state cannot drive winds from the disk, and so Springel \& Hernquist (2003) implement an {\it ad hoc} wind model, which expels gas from the disk at a fixed velocity. 

This relatively stiff equation of state suppresses the Toomre instability in the disk and produces a star formation rate which agrees with the Kennicutt (1989) star formation law.  In addition, it seems to produce disks with rotation curves in agreement with observations and a correct Tully-Fisher relation (Robertson \etal 2004), at least for low-mass galaxies. One criticism of this model is that it predicts an extremely high thermal pressure for typical disk conditions -- for example the predicted pressure at 1 $cm^{-3}$ is nearly two orders of magnitude higher than suggested by our current understanding of the ISM thermal pressure (e.g. Boulares \& Cox  1990; Heiles 1989), which is probably small compared to the turbulent and/or magnetic pressures.


\section{Models of the Interstellar Medium}
\label{sec:ism_models}

One way to select an appropriate ``sub-grid" model is to turn to observations and models of the local ISM.  Unfortunately, the Milky Way's ISM is still incompletely understood, but models are becoming more sophisticated.  Here we present an incomplete review of some models, focusing on numerical models.  A good model of the ISM in a galaxy would simulate the entire disk in order to resolve large-scale gravitational features, and go down to scales of at least 1 pc, in order to model the formation of molecular clouds (even this, of course, is insufficient to model the formation of molecular cores and star formation, but at least suffices to follow the formation of the majority of the molecular clouds).  In addition, the model would include MHD and radiative transfer effects, as well as energy and mass input from massive stars.  What can be accomplished with current technology is, naturally, somewhat less.

\subsection{Global Disk Models}

One simplification is to reduce the dimensionality of the problem, as done by Wada \& Norman (1999, 2001), who obtained 2 pc resolution in a global disk model in two dimensions.  They found that the medium quickly fragmented into small knots with interconnected filaments, forming a three phase medium, with gas ranging from 10 K to $10^6$ K.  The thermal pressure was varied, but did not form a monotonic relation between pressure and density.  On the other hand, by defining an effective pressure which was the thermal plus turbulent pressure, they did find a relation between density and effective pressure which was mildly steeper than isothermal.  By defining a Toomre $Q$ parameter using this effective pressure, Wada \& Norman found that the disk became marginally stable ($Q \approx 2$) over most of the radius of the disk.

While promising, it is known that two-dimensional and three dimensional turbulence have different properties, but to explore global disk models in three dimensions means reducing the resolution.  This approach was taken by Tasker \& Bryan (2006), using the same AMR code described earlier.  In that work, a global Milky-Way like model is simulated with resolution up to 25 pc, good enough to see the large molecular cloud formation.  This simulation also found a natural three-phase medium, and, in addition, reproduced the Kennicutt-Schmidt relation for star formation as a function of surface density ($\Sigma_{\rm SFR} \propto \Sigma_{\rm gas}^{1.5}$), assuming that star formation only occurred in dense molecular clouds with densities larger than about $10^3$ cm$^{-3}$.  Interestingly, this result was found whether or not feedback from SN was included, although such feedback did drive a galactic fountain from the surface of the disk.

\subsection{Local ISM Models}

While current global disk models are limited by resolution, it is possible to simulate a local region of the ISM at high resolution.  This has been done by a number of authors (e.g. Korpi \etal 1999; de Avillez \& Breitschwerdt 2004), and the result is always a multi-phase medium.  Such simulations resolve the supernovae feedback and find that the blast waves drive a turbulent medium.  Recently, Joung \& Mac Low (2006) simulated a 0.5 x 0.5 kpc$^2$ section of the disk, using AMR to model the height structure up to 10 kpc away from the disk.   They found a turbulent multi-phase medium with a characteristic scale of turbulent driving around 100 pc (in particular they found that 90\% of the energy was on scales below 200 pc).  Extensions of models like this to MHD (de Avillez \& Breitschwerdt 2005) find that for reasonable magnetic field choices, the magnetic pressure only dominates in the coldest gas ($T < 100$ K).  The also find that turbulent pressures are at least a few times the thermal pressure except in the hottest part of the medium.


\section{New directions in subgrid models}

As we have seen, the ISM models suggest that turbulent pressure is more important than either thermal or magnetic pressures in most ISM conditions.  This suggests that cosmological sub-grid models should include a term for the turbulent pressure, rather than focusing on thermal forces.  Here we suggest a simple extension that would permit such a description in a fluid equation.  Essentially, we add an additional energy term to describe the turbulent energy on scales not resolved by the simulation.  Since most cosmological simulations only resolve scales larger than a few hundred pc, and most of the turbulent energy is on scales less than this, there is some hope that such a separation of scales model could work.  The pressure would be given by
$$
P = (\gamma - 1)(e + q) \rho
$$
where $e$ is the traditional thermal energy (per unit mass), while $q$ is the specific turbulent energy density.  We require another energy equation, which should be like the one for the thermal energy but without any radiative cooling terms, and should also include a source term reflecting the turbulent energy input from supernovae, and a sink term which corresponds to the decay of turbulence (Mac Low 1999):
$$
\frac{dq}{dt} = \epsilon_1 \dot{\rho}_{\rm SN} - \epsilon_2 \frac{q^{3/2}}{L}
$$
The turbulent energy is assumed to decay at rate proportional to $q/t_{\rm cross}$ where $t_{\rm cross} = L/q^{1/2}$ is the crossing time at the turbulent driving scale ($L \approx 100$ pc).  The two parameters $\epsilon_1$ and $\epsilon_2$ would be selected to match local ISM simulations.

Such a model would naturally include the effect of turbulence in large-scale simulations, at least as far as we understand it from small-scale models.  It would treat turbulence as strictly a pressure term, although small-scale effects on the star formation rate could, in principle, be included in a statistical way.  The advantage is that we would be prescribing the sub-grid model from what we know (bottom-up) rather than what we need in order to resolve current problems in cosmological simulations (top-down).  


\section{Conclusions}

It appears that we are on the cusp of being able to create realistic galaxies from (nearly) first-principle simulations.  Assuming this can be done, we will be able to answer many long-standing questions about the nature of dark matter and the origin of the Tully-Fisher relation.  However, before this can be done, we must gain a more complete understanding of how the ISM affects galaxy formation.  As this review has shown, we are not quite there.

I thank the organizers for a very enjoyable and enlightening conference and my collaborators for allowing me to share our work.



\begin{thebibliography}{99}


\bibitem[Abadi et al.(2003)]{2003ApJ...591..499A} Abadi, M.~G., Navarro, 
J.~F., Steinmetz, M., \& Eke, V.~R.\ 2003, \apj, 591, 499 


\bibitem[Abadi et al.(2003)]{2003ApJ...597...21A} Abadi, M.~G., Navarro, 
J.~F., Steinmetz, M., \& Eke, V.~R.\ 2003, \apj, 597, 21 


\bibitem[Barnes \& Hernquist(1992)]{1992Natur.360..715B} Barnes, J.~E., \& 
Hernquist, L.\ 1992, \nat, 360, 715 


\bibitem[Boulares \& Cox(1990)]{1990ApJ...365..544B} Boulares, A., \& Cox, 
D.~P.\ 1990, \apj, 365, 544 


\bibitem[Brook et al.(2004)]{2004ApJ...612..894B} Brook, C.~B., Kawata, D., 
Gibson, B.~K., \& Freeman, K.~C.\ 2004, \apj, 612, 894 


\bibitem[Bryan \& Norman(1997)]{1997ASPC..123..363B} Bryan, G.~L., \& 
Norman, M.~L.\ 1997, ASP Conf.~Ser.~123: Computational Astrophysics; 12th 
Kingston Meeting on Theoretical Astrophysics, 123, 363 

\bibitem[Bryan(1999)]{bryan1999} Bryan, G. L., 1999, Comput. Sci. Eng., 1, 46


\bibitem[Chung et al.(2002)]{2002AJ....123.2387C} Chung, A., van Gorkom, 
J.~H., O'Neil, K., \& Bothun, G.~D.\ 2002, \aj, 123, 2387 


\bibitem[de Avillez \& Breitschwerdt(2005)]{2005A&A...436..585D} de 
Avillez, M.~A., \& Breitschwerdt, D.\ 2005, \aap, 436, 585 


\bibitem[de Avillez \& Breitschwerdt(2004)]{2004Ap&SS.289..479D} de 
Avillez, M., \& Breitschwerdt, D.\ 2004, \apss, 289, 479 


\bibitem[de Blok(2004)]{2004IAUS..220...69D} de Blok, W.~J.~G.\ 2004, IAU 
Symposium, 220, 69 


\bibitem[de Blok et al.(2003)]{2003MNRAS.340..657D} de Blok, W.~J.~G., 
Bosma, A., \& McGaugh, S.\ 2003, \mnras, 340, 657 


\bibitem[Diemand et al.(2004)]{2004MNRAS.353..624D} Diemand, J., Moore, B., 
\& Stadel, J.\ 2004, \mnras, 353, 624 


\bibitem[Eisenstein et al.(2005)]{2005ApJ...633..560E} Eisenstein, D.~J., 
et al.\ 2005, \apj, 633, 560 


\bibitem[Fall \& Efstathiou(1980)]{1980MNRAS.193..189F} Fall, S.~M., \& 
Efstathiou, G.\ 1980, \mnras, 193, 189 


\bibitem[Fryxell et al.(2000)]{2000ApJS..131..273F} Fryxell, B., et al.\ 
2000, \apjs, 131, 273 


\bibitem[Fukushige et al.(2004)]{2004ApJ...606..625F} Fukushige, T., Kawai, 
A., \& Makino, J.\ 2004, \apj, 606, 625 


\bibitem[Gerritsen(1997)]{1997PhDT........19G} Gerritsen, J.~P.~E.\ 1997, 
Ph.D.~Thesis,  


\bibitem[Gnedin \& Bertschinger(1996)]{1996ApJ...470..115G} Gnedin, N.~Y., 
\& Bertschinger, E.\ 1996, \apj, 470, 115 


\bibitem[Governato et al.(2006)]{2006astro.ph..2351G} Governato, F., 
Willman, B., Mayer, L., Brooks, A., Stinson, G., Valenzuela, O., Wadsley, 
J., \& Quinn, T.\ 2006, ArXiv Astrophysics e-prints, arXiv:astro-ph/0602351 


\bibitem[Hayashi et al.(2004)]{2004MNRAS.355..794H} Hayashi, E., et al.\ 
2004, \mnras, 355, 794 


\bibitem[Heiles(1989)]{1989ApJ...336..808H} Heiles, C.\ 1989, \apj, 336, 
808 


\bibitem[Joung \& Mac Low(2006)]{2006astro.ph..1005J} Joung, M.~K.~R., \& 
Mac Low, M.-M.\ 2006, ArXiv Astrophysics e-prints, arXiv:astro-ph/0601005 


\bibitem[Kaufmann et al.(2006)]{2006astro.ph..1115K} Kaufmann, T., Mayer, 
L., Wadsley, J., Stadel, J., \& Moore, B.\ 2006a, ArXiv Astrophysics 
e-prints, arXiv:astro-ph/0601115 


\bibitem[Kaufmann et al.(2006)]{2006MNRAS.370.1612K} Kaufmann, T., Mayer, 
L., Wadsley, J., Stadel, J., \& Moore, B.\ 2006b, \mnras, 370, 1612 


\bibitem[Kennicutt(1989)]{1989ApJ...344..685K} Kennicutt, R.~C., Jr.\ 1989, 
\apj, 344, 685 


\bibitem[Klypin et al.(1999)]{1999ApJ...522...82K} Klypin, A., Kravtsov, 
A.~V., Valenzuela, O., \& Prada, F.\ 1999, \apj, 522, 82 


\bibitem[Korpi et al.(1999)]{1999ApJ...514L..99K} Korpi, M.~J., 
Brandenburg, A., Shukurov, A., Tuominen, I., \& Nordlund, {\AA}.\ 1999, 
\apjl, 514, L99 


\bibitem[Li et al.(2005)]{2005ApJ...626..823L} Li, Y., Mac Low, M.-M., \& 
Klessen, R.~S.\ 2005, \apj, 626, 823 

\bibitem[Mac Low(1999)]{1999ApJ...524..169M} Mac Low, M.-M.\ 1999, \apj, 
524, 169 

\bibitem[Mandelbaum et al.(2006)]{2006MNRAS.368..715M} Mandelbaum, R., 
Seljak, U., Kauffmann, G., Hirata, C.~M., \& Brinkmann, J.\ 2006, \mnras, 
368, 715 


\bibitem[Marri \& White(2003)]{2003MNRAS.345..561M} Marri, S., \& White, 
S.~D.~M.\ 2003, \mnras, 345, 561 


\bibitem[Moore et al.(1999)]{1999ApJ...524L..19M} Moore, B., Ghigna, S., 
Governato, F., Lake, G., Quinn, T., Stadel, J., \& Tozzi, P.\ 1999, \apjl, 
524, L19 


\bibitem[Moore et al.(2004)]{2004MNRAS.354..522M} Moore, B., Kazantzidis, 
S., Diemand, J., \& Stadel, J.\ 2004, \mnras, 354, 522 


\bibitem[Navarro et al.(2004)]{2004MNRAS.349.1039N} Navarro, J.~F., et al.\ 
2004, \mnras, 349, 1039 


\bibitem[Navarro \& Benz(1991)]{1991ApJ...380..320N} Navarro, J.~F., \& 
Benz, W.\ 1991, \apj, 380, 320 


\bibitem[Navarro \& White(1994)]{1994MNRAS.267..401N} Navarro, J.~F., \& 
White, S.~D.~M.\ 1994, \mnras, 267, 401 


\bibitem[Norman \& Bryan(1999)]{1999numa.conf...19N} Norman, M.~L., \& 
Bryan, G.~L.\ 1999, ASSL Vol.~240: Numerical Astrophysics, 19 


\bibitem[Okamoto et al.(2005)]{2005MNRAS.363.1299O} Okamoto, T., Eke, 
V.~R., Frenk, C.~S., \& Jenkins, A.\ 2005, \mnras, 363, 1299 

\bibitem[O'Shea et al.(2004)]{oshea2004} O'Shea, B. W., Bryan, G. L., Bordner, J., Norman, M. L., Abel, T., Harkness, R., \& Kritsuk, A., 2004, astro-ph/0403044


\bibitem[Power et al.(2003)]{2003MNRAS.338...14P} Power, C., Navarro, 
J.~F., Jenkins, A., Frenk, C.~S., White, S.~D.~M., Springel, V., Stadel, 
J., \& Quinn, T.\ 2003, \mnras, 338, 14 


\bibitem[Robertson et al.(2004)]{2004ApJ...606...32R} Robertson, B., 
Yoshida, N., Springel, V., \& Hernquist, L.\ 2004, \apj, 606, 32 


\bibitem[Semelin \& Combes(2002)]{2002A&A...388..826S} Semelin, B., \& 
Combes, F.\ 2002, \aap, 388, 826 


\bibitem[Slyz et al.(2005)]{2005MNRAS.356..737S} Slyz, A.~D., Devriendt, 
J.~E.~G., Bryan, G., \& Silk, J.\ 2005, \mnras, 356, 737 


\bibitem[Sommer-Larsen et al.(2003)]{2003ApJ...596...47S} Sommer-Larsen, 
J., G{\"o}tz, M., \& Portinari, L.\ 2003, \apj, 596, 47 


\bibitem[Springel et al.(2005)]{2005MNRAS.361..776S} Springel, V., Di 
Matteo, T., \& Hernquist, L.\ 2005, \mnras, 361, 776 


\bibitem[Springel \& Hernquist(2003)]{2003MNRAS.339..289S} Springel, V., \& 
Hernquist, L.\ 2003, \mnras, 339, 289 


\bibitem[Steinmetz \& Muller(1995)]{1995MNRAS.276..549S} Steinmetz, M., \& 
Muller, E.\ 1995, \mnras, 276, 549 


\bibitem[Stinson et al.(2006)]{2006astro.ph..2350S} Stinson, G., Seth, A., 
Katz, N., Wadsley, J., Governato, F., \& Quinn, T.\ 2006, ArXiv 
Astrophysics e-prints, arXiv:astro-ph/0602350 


\bibitem[Swaters et al.(2003)]{2003ApJ...587L..19S} Swaters, R.~A., 
Verheijen, M.~A.~W., Bershady, M.~A., \& Andersen, D.~R.\ 2003, \apjl, 587, 
L19 


\bibitem[Tasker \& Bryan(2006)]{2006ApJ...641..878T} Tasker, E.~J., \& 
Bryan, G.~L.\ 2006, \apj, 641, 878 


\bibitem[Thacker \& Couchman(2001)]{2001ApJ...555L..17T} Thacker, R.~J., \& 
Couchman, H.~M.~P.\ 2001, \apjl, 555, L17 


\bibitem[van den Bosch et al.(2000)]{2000AJ....119.1579V} van den Bosch, 
F.~C., Robertson, B.~E., Dalcanton, J.~J., \& de Blok, W.~J.~G.\ 2000, \aj, 
119, 1579 


\bibitem[Wada \& Norman(1999)]{1999ApJ...516L..13W} Wada, K., \& Norman, 
C.~A.\ 1999, \apjl, 516, L13 


\bibitem[Wada \& Norman(2001)]{2001ApJ...547..172W} Wada, K., \& Norman, 
C.~A.\ 2001, \apj, 547, 172 


\bibitem[Weil et al.(1998)]{1998MNRAS.300..773W} Weil, M.~L., Eke, V.~R., 
\& Efstathiou, G.\ 1998, \mnras, 300, 773 


\bibitem[Yepes et al.(1997)]{1997MNRAS.284..235Y} Yepes, G., Kates, R., 
Khokhlov, A., \& Klypin, A.\ 1997, \mnras, 284, 235 




\end{thebibliography}
\end{document}